\documentclass[aip,graphicx,amsmath,reprint]{revtex4-1}
\usepackage{graphicx}
\usepackage{dcolumn}
\usepackage{bm}

\usepackage[utf8]{inputenc}
\usepackage[T1]{fontenc}
\usepackage{mathptmx}
\usepackage{etoolbox}
\usepackage{color}
\usepackage{ulem}

\makeatletter
\def\@email#1#2{%
 \endgroup
 \patchcmd{\titleblock@produce}
  {\frontmatter@RRAPformat}
  {\frontmatter@RRAPformat{\produce@RRAP{*#1\href{mailto:#2}{#2}}}\frontmatter@RRAPformat}
  {}{}
}%
\makeatother
\begin{document}

\title{Nonlinear anisotropic equilibrium reconstruction in  axisymmetric magnetic mirrors} 



\author{S. J. Frank}
\email{sfrank@realtafusion.com}
\affiliation{Realta Fusion, Madison, WI}

\author{I. Agarwal}
\affiliation{Realta Fusion, Madison, WI}

\author{J. K. Anderson}
\affiliation{University of Wisconsin-Madison, Madison, WI}

\author{B. Biswas}
\affiliation{Realta Fusion, Madison, WI}

\author{E. Claveau}
\affiliation{Realta Fusion, Madison, WI}

\author{D. Endrizzi}
\affiliation{Realta Fusion, Madison, WI}

\author{C. Everson}
\affiliation{Realta Fusion, Madison, WI}

\author{R. W. Harvey}
\affiliation{CompX, Del Mar, CA}

\author{S. Murdock}
\affiliation{University of Wisconsin-Madison, Madison, WI}

\author{Yu. V. Petrov}
\affiliation{CompX, Del Mar, CA}

\author{J. Pizzo}
\affiliation{University of Wisconsin-Madison, Madison, WI}

\author{T. Qian}
\affiliation{Princeton University, Princeton, NJ}

\author{K. Sanwalka}
\affiliation{University of Wisconsin-Madison, Madison, WI}

\author{K. Shih}
\affiliation{Realta Fusion, Madison, WI}

\author{D.A. Sutherland}
\affiliation{Realta Fusion, Madison, WI}

\author{A. Tran}
\affiliation{University of Wisconsin-Madison, Madison, WI}

\author{J. Viola}
\affiliation{Massachusetts Institute of Technology, Cambridge, MA}

\author{D. Yakovlev}
\affiliation{University of Wisconsin-Madison, Madison, WI}

\author{M. Yu}
\affiliation{University of Wisconsin-Madison, Madison, WI}

\author{C. B. Forest}
\affiliation{University of Wisconsin-Madison, Madison, WI}


\date{\today}

\begin{abstract}
Magnetic equilibrium reconstruction is a crucial simulation capability for interpreting diagnostic measurements of experimental plasmas. Equilibrium reconstruction has mostly been applied to systems with isotropic pressure and relatively low plasma $\beta = 2\mu_0p/B^2$. This work extends nonlinear equilibrium reconstruction to high-$\beta$ plasmas with anisotropic pressure and applies it to the Wisconsin High Temperature Superconducting Axisymmetric Magnetic Mirror experiments to infer the presence of sloshing ions. A novel basis set for the plasma profiles and machine learning algorithm using scalable constrained Bayesian optimization allow accurate nonlinear reconstructions with uncertainty quantification to be made more quickly with fewer experimental diagnostics and improves the robustness of reconstructions at high $\beta$. In addition to WHAM and other mirrors, such reconstruction techniques are potentially attractive in high-performance devices with constrained diagnostic capabilities such as fusion power plants.
\end{abstract}


\maketitle 

\section{Introduction and background}\label{sec:intro}

Magnetic equilibrium reconstruction is a fundamental experimental capability originally developed for tokamaks \cite{Lao1985,Lao1990} and has been extended to several different types of fusion devices.\cite{Quon1985,Anderson2004,Hanson2009,Dettrick2021} To perform an axisymmetric anisotropic equilibrium reconstruction, we solve the Grad-Shafranov (GS) equation:\cite{Grad1967}
\begin{equation}\label{eq:GS}
        \Delta^*\psi = - \mu_0 R J_\phi,
\end{equation}
where $\psi$ is the magnetic flux normalized by $2\pi$, $\mu_0$ is the vacuum magnetic permeability, and $\vec J$ is the current density, $\Delta^* = \frac{\partial^2}{\partial R^2} -\frac{1}{R}\frac{\partial}{\partial R} + \frac{\partial ^2}{\partial Z^2}$, and we have assumed a cylindrical $(R,\phi,Z)$ coordinate system. The value of $J_\phi$ in the plasma may be obtained by utilizing the equation for magnetohydrodynamic (MHD) force balance:
\begin{equation}\label{eq:forceBal}
    \nabla \cdot \underline{\underline{P}} = \vec{J} \times \vec{B},
\end{equation}
where $\underline{\underline{P}}$ is the pressure tensor and $\vec B$ is the magnetic field. We assume an anisotropic pressure tensor with form $\underline{\underline{P}} = p_\perp \underline{\underline{I}} - (p_\perp - p_\parallel)\hat b \hat b$, in which $(\parallel, \perp)$ denote the direction of the pressure relative to the magnetic field direction $\hat b = \vec B/B$.  In the anisotropic case, $p_{\parallel,\perp}$ are functions of $\psi$ and $B$. We exclude flows here. The Mach number $M=v^2/v_{cs}^2 \approx v^2m_i/T_i$, where $v$ is the flow velocity, $m_i$ is the ion mass, and $T_i$ is the ion temperature. The Alfv\'enic Mach number $M_A^2 = v^2/v_A^2 = \mu_0 \rho v^2/B^2$, where $v_A$ is the Alfv\'en speed and $\rho$ is the mass density. In the WHAM plasmas under consideration here, perpendicular flows have velocities $v\sim10~\mathrm{km/s}$, the sound speed $v_{cs}\sim100~\mathrm{km/s}$, and the Alfv\'en speed $v_A\sim1000~\mathrm{km/s}$. Therefore, $M_A^2\ll M^2\ll1$. The value of $M^2$ can become appreciable for parallel flows in the expanders and close to the magnetic mirror throat. However, in these regions $M_A^2$ is often nearly vanishing due to large $B$ and low plasma densities. Thus, there are only very small perturbations to the vacuum fields. The force balance above may be written in terms of its perpendicular:
\begin{equation}\label{eq:forceBalPrp}
    \nabla_\perp\left(\frac{B^2}{2\mu_0} + p_\perp \right) = \left(\frac{B^2}{\mu_0} + p_\perp - p_\parallel \right)\vec \kappa,
\end{equation}
and parallel:
\begin{equation}\label{eq:forceBalPar}
    p_\perp = - B^2 \frac{\partial}{\partial B}\Big(\frac{p_\parallel}{B}\Big),
\end{equation}
components, where $\vec \kappa = \hat b\cdot\nabla\hat b$ is the magnetic curvature. For a valid equilibrium to exist, conditions:
\begin{equation}\label{eq:firehose}
    p_\parallel - p_\perp - \frac{B^2}{\mu_0} < 0
\end{equation}
\begin{equation}\label{eq:mirror}
    B\frac{\partial p_\perp}{\partial B} -\frac{B^2}{\mu_0} < 0,
\end{equation}
must be satisfied. The first condition (\ref{eq:firehose}) is the firehose instability stability condition, and the second condition (\ref{eq:mirror}) is the mirror instability stability condition.\cite{Ryutov2011}

Rewriting the perpendicular force balance as an equation for current density allows us to solve (\ref{eq:GS}):
\begin{equation}
    \vec J_\phi \equiv \vec J_\perp = \frac{\vec B}{B^2} \times \left[\nabla_\perp p_\perp + (p_\parallel - p_\perp) \vec{\kappa} \right].
\end{equation}
To perform a kinetic equilibrium reconstruction, we use some profiles of the distribution function $f(\psi,B,C_n)$ with parameterization constants $C_n$ and integrate them to obtain $p_{\parallel,\perp}(\psi,B,C_n)$. This provides a solution for $\psi$ and $B$ for some given $f$ profiles. The simulated equilibrium parameters are compared to diagnostic measurements for varying values of $C_n$, and a minimum error solution can be obtained through techniques such as $\chi^2$ minimization between the simulation results and diagnostic measurements \cite{Quon1985,Anderson2004} or the EFIT algorithm.\cite{Lao1985} 

In this work, we propose a new method for machine-learning accelerated nonlinear magnetic equilibrium reconstruction with automatic uncertainty quantification in axisymmetric plasmas with anisotropic pressure and high $\beta = 2\mu_0 p/B^2$ using a novel kinetic basis function. This yields self-consistent anisotropic pressure and density profiles that allow us to constrain equilibrium reconstructions with a wider array of measurements and provide accurate measures of the plasma $\beta$, total stored energy $W_\mathrm{tot}$ and the average ion energy $\langle E_i\rangle$, which are key parameters for assessing plasma performance metrics. We describe the \texttt{Pleiades} code used for the reconstructions, the construction of the kinetic basis functions, the machine-learning fitting technique, and verify our reconstruction technique with synthetic data in Section~\ref{sec:code}. We then apply our equilibrium reconstructions to the Wisconsin High Temperature Superconducting Axisymmetric Mirror (WHAM) \cite{Endrizzi2023} experiments described in Fujii \textit{et al.}\cite{Fujii2025} in Section~\ref{sec:exp}.

While this work uses \texttt{Pleiades}\cite{Peterson2019} to perform reconstructions, the techniques described here for kinetic reconstruction and machine-learning enhanced nonlinear equilibrium fitting are generalizable. It should be straightforward to adapt them to any other Grad-Shafranov solver.

\section{Equilibrium reconstruction with \texttt{Pleiades}} \label{sec:code}

\subsection{The \texttt{Pleiades} equilibrium solver}
\texttt{Pleiades} is a Green's function based free-boundary Grad-Shafranov equilibrium solver written in object oriented Python with some OpenMP accelerated Fortran. \texttt{Pleiades} was originally developed as a magnetostatics code\cite{Peterson2019} and extended to be a fully-fledged equilibrium solver in later work.\cite{Endrizzi2023,Forest2024,Frank2025} To solve (\ref{eq:GS}), \texttt{Pleiades} uses a Green's function solution to the free-boundary GS equation:
\begin{equation}\label{eq:GS_greens}
    \psi(\vec x) = \int_{-\ell}^\ell\int_0^a \mathcal{G}(\vec x, \vec x^\prime) J_\phi (\vec x^\prime ) dRdZ,
\end{equation}
where $\ell$ is the length of the plasma referenced to zero at the midplane, $a$ is the plasma radius, and the Green's function in cylindrical coordinates is:
\begin{eqnarray}
    \mathcal{G}(R,Z,R^\prime ,Z^\prime) = \frac{\mu_0}{2\pi} \frac{\sqrt{R^\prime R}}{k}\left[\left(2 - k^2 \right) K(k) - 2E(k)\right].
\end{eqnarray}
where $K$ and $E$ are complete elliptic integrals of the first and second kind respectively and $k^2 = 4 R R^\prime / [(R  + R^\prime)^2 + (Z-Z^\prime)^2]$. We discretize (\ref{eq:GS_greens}) on an axisymmetric $(R,Z)$ grid. We solve for $\psi$ at each $(R, Z)$ point of interest at $i$ using the $J_\phi$ resultant from currents in the plasma at $j$, magnets at $k$, and perfectly conducting boundaries at $l$ at spatial locations $(R^\prime, Z^\prime)_{\{j,k,l \}}$. Using this discretization (\ref{eq:GS_greens}) becomes:
\begin{eqnarray}
    \psi_{i} = \sum_j \mathcal{G}_{ij}J^{\mathrm{plas}}_{\phi, j} \Delta R_j^\prime \Delta Z_j^\prime + \sum_k \mathcal{G}_{ik} J^\mathrm{mag}_{\phi, k} \Delta R_k^\prime \Delta Z_k^\prime \nonumber \\
       + \sum_l \mathcal{G}_{il} J^\mathrm{BC}_{\phi, l} \Delta R_l^\prime \Delta Z_l^\prime,
\end{eqnarray}
where $\Delta R \Delta Z$ is the area of a grid cell and matrices $\mathcal{G}_{i\{j,k,l\}}$ are the Green's function matrices relating magnetic flux to the current sources in the plasma, magnets, and any perfectly conducting boundaries respectively. The values of $J_\phi^\mathrm{mag}$ are obtained from the experimental parameters of the magnets, and values of the diamagnetic plasma current $J_\phi^\mathrm{plas}$ are obtained using:
\begin{eqnarray}\label{eq:j_plas}
    J_\phi^{plas} = \frac{1}{B}\Bigg[B_Z \frac{\partial p_\perp}{\partial R} - B_R \frac{\partial p_\perp}{\partial Z} \nonumber \\      + \frac{p_\parallel - p_\perp}{B}  \left(\kappa_Z\frac{\partial B}{\partial R} - \kappa_R \frac{\partial B}{\partial Z}\right)\Bigg].
\end{eqnarray}
To calculate (\ref{eq:j_plas}), \texttt{Pleiades} uses pressure profiles obtained from either a transport calculation performed with a code like \texttt{CQL3D-m} \cite{Harvey2016} or with model profiles. In either case, profiles must satisfy parallel GS equilibrium condition (\ref{eq:forceBalPar}).

To compute an equilibrium, \texttt{Pleiades} evaluates the vacuum values of $\psi$ and $B$, interpolates $p_{\parallel,\perp} (\psi,B)$ onto the vacuum $\psi$ and $B$ profiles, calculates the plasma contribution to $\psi$ and $B$, re-interpolates $p_{\parallel,\perp} (\psi,B)$ onto the updated $\psi$ and $B$ profiles including the plasma contribution, and then once again recalculates the plasma contributions to $\psi$ and $B$. This process is repeated until the relative change in $\psi$ of the current iteration versus the previous iteration drops below a user input tolerance level. For ease of numerical implementation, we generally represent $p_{\parallel,\perp} (\psi,b)$, where $b = B/B_0$ is the magnetic field normalized by the midplane field $B_0$. 

If a perfectly conducting wall boundary condition $\Delta \psi^\mathrm{BC} = 0$ is present, after the calculation of the plasma response in the iterative GS solve, an additional step to calculate $J^\mathrm{BC}_\phi$ can be taken. This is done by solving equation:
\begin{equation}
       -\Delta\psi_n^\mathrm{BC} = \sum_l \mathcal{G}_{nl} J_{\phi,l}^\mathrm{BC} \Delta R_l \Delta Z_l,
\end{equation}
where $\psi_n^\mathrm{BC}$ is the flux at $n$ $(R,Z)$ points on the perfectly conducting boundary. Using singular value decomposition (SVD) to obtain a pseudo-inverse of the Green's function relating currents in the perfect conductor to the surface of the perfect conductor, the above equation can be inverted and solved to obtain a solution for $J_\phi^\mathrm{BC}$ which approximately enforces the boundary condition. The contribution to the total flux from this boundary condition is then included in the iterative solution to the GS equation and updated on each iteration. As the Green's function SVD can be precomputed, enforcing this boundary condition is very fast. In WHAM, the vacuum vessel wall is distant from the plasma and the flux loops.\cite{Endrizzi2023} In the limit of a symmetric perfectly conducting wall, \texttt{Pleiades} simulations indicate wall currents have a minimal effect on the diamagnetic flux measurements in present WHAM experiments.

\texttt{Pleiades} equilibrium reconstructions use fixed $(R,Z)$ field grids and fixed $(R^\prime,Z^\prime)$ current grids so that Green's functions can be calculated just once prior to the iterative GS solve. After this precomputation of the Green's functions, the iterative solution method requires only fast matrix multiplications to be performed rather than costly recomputation of the Green's functions. 
\subsection{A semi-analytic kinetic basis for anisotropic mirror pressure profiles}
\begin{figure}
    \centering
    \includegraphics[width=\linewidth]{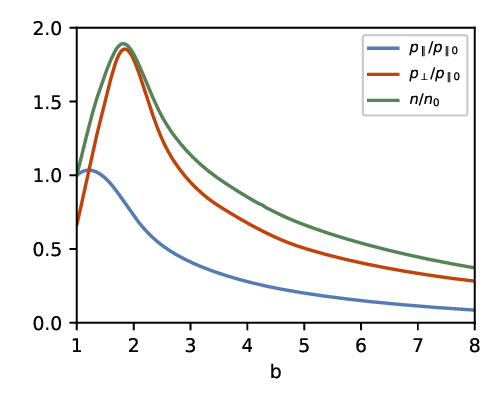}
    \caption{Normalized parallel pressure $p_\parallel/p_{\parallel 0}$, perpendicular pressure $p_\perp/p_{\parallel 0}$, and density $n/n_0$ for $m_i=2~\mathrm{u}$, $R_m = 20$, $T_e=1~\mathrm{keV}$, $Z_\mathrm{eff} = 1.0$, $E_\mathrm{NBI} = 25~\mathrm{keV}$, and $\theta_\mathrm{NBI} = 45^\circ$.}
    \label{fig:egedal_profs}
\end{figure}
We will derive an analytic form for the distribution function and resultant anisotropic equilibrium pressure and density profiles in a neutral beam injection (NBI) fueled mirror plasma with sloshing ions based on a modification of the work in Bilikmen \textit{et al.}\cite{Bilikmen1997} and Egedal \textit{et al.}\cite{Egedal2022} To begin, the total pressure profiles may be described:
\begin{eqnarray}
    \label{eq:p_par_prof} p_\parallel (\psi,b) =& p_\mathrm{GD}(\psi)  + (n_\mathrm{GD}(\psi) + n_\mathrm{hot}(\psi,b))T_e(\psi) \nonumber \\ & + p_\mathrm{hot,\parallel}(\psi,b)\\
    p_\perp(\psi,b) =& p_\mathrm{GD}(\psi) + (n_\mathrm{GD}(\psi) + n_\mathrm{hot}(\psi,b))T_e(\psi) \nonumber\\ &+ p_\mathrm{hot,\perp}(\psi,b)\\
    \label{eq:p_prof} p(\psi, b) =& (p_\parallel+2p_\perp)/3, 
\end{eqnarray}
where $p_\mathrm{GD}$ is the cool Maxwellian gas dynamic pressure,  $p_\mathrm{hot,(\parallel ,\perp)}(\psi,b)$ are the anisotropic pressures from the hot ions, $n_{GD}(\psi)$ is the gas dynamic plasma density, $n_\mathrm{hot}(\psi,b)$ is the hot ion density, and $T_e$ is the electron temperature. In principle, hot electron distributions from electron cyclotron heating (ECH) can also generate sloshing electron pressure profiles. The pressure contribution from these electrons has not been included in the pressure functions here (\ref{eq:p_par_prof}-\ref{eq:p_prof}), but will be investigated in future work. We can construct $p_\mathrm{hot}$ from model distribution functions through integration:
\begin{subequations}
\begin{eqnarray}
    p_{\parallel, \mathrm{hot}}(\psi,b) =& 2\pi m_i\mathcal{A}(\psi)\int_{0}^{v_0} dv \int_0^\pi d\theta \nonumber\\ &\times  [f_\mathrm{hot}(v,\theta,b)/\tau_s] v^4 \cos^2\theta \sin \theta   \\
    p_{\perp, \mathrm{hot}}(\psi,b) =& \pi m_i \mathcal{A}(\psi) \int_{0}^{v_0} dv \int_0^\pi d\theta \nonumber\\  &\times  [f_\mathrm{hot}(v,\theta,b)/\tau_s] v^4 \sin^3 \theta,   
\end{eqnarray}
\end{subequations}
where $v$ is the velocity, $\theta$ is the pitch angle, $v_0$ is the neutral beam injection velocity, $m_i$ is the ion mass, $\mathcal{A}(\psi)$ is the radial pressure profile multiplied by the fast ion slowing down time $\tau_s$ (multiplying the pressure profile by the slowing down time simplifies normalizations substantially), and $f_\mathrm{hot}(v,\theta)$ is the hot ion distribution function. To generate axial pressure profiles, we must pick some distribution basis $f_\mathrm{hot}(v,\theta)$ which can be evaluated quickly but is detailed enough to provide an accurate approximation of mirror confinement dynamics. A convenient form is found in Egedal, \textit{et al.} \cite{Egedal2022}:
\begin{equation}\label{eq:f_egedal}
    f_\mathrm{hot}(v,\theta) = \frac{\tau_s}{v^3+v_c^3}\sum_jS_jM_{\lambda_j}(\theta)u^{\lambda_j},
\end{equation}
Details on the terms in this equation and its derivation may be found in Appendix~\ref{ap:FP_sol}. The NBI source in WHAM may be modeled with:
\begin{equation}
    S_\mathrm{NBI} = \delta(v-v_\mathrm{NBI}) \delta(\theta - \theta_\mathrm{NBI}),
\end{equation}
where $v_\mathrm{NBI} = \sqrt{2E_\mathrm{NBI}/m_i}$ is the neutral beam velocity and $\theta_\mathrm{NBI}$ is the neutral beam injection angle. If (\ref{eq:f_egedal}) is taken to be the midplane distribution function, it is possible to find the off-midplane distribution at a given value of $b$ by integrating: 
\begin{subequations}
\label{eq:hot_i_basis}
\begin{eqnarray}
    p_{\parallel, \mathrm{hot}}(\psi,b) =& 4\pi m_i\mathcal{A}(\psi)\int_{0}^{v_0} dv \int_0^{\xi_\mathrm{PT,loc}(b)} d\xi \nonumber \\  &\times v^4 \xi^2 [f_\mathrm{hot}(v,\xi_0(\xi,b))/\tau_s] \\
    p_{\perp, \mathrm{hot}}(\psi,b) =& 2\pi m_i \mathcal{A}(\psi) \int_{v_c}^{v_0} dv\nonumber\int_0^{\xi_\mathrm{PT,loc}(b)} d\xi  \\ &\times v^4 (1-\xi^2) [f_\mathrm{hot}(v,\xi_0(\xi,b))/\tau_s],   
\end{eqnarray}
\end{subequations}
where $\xi = v_\parallel/v$ is a pitch angle like coordinate, $\xi_\mathrm{TP,loc} = \sqrt{1-1/R_{m,\mathrm{loc}}}$ is the local trapped-passing boundary for $R_{m,\mathrm{loc}} = B_m/B = R_m/b$ with $B_m$ equal to the magnetic field at the magnetic mirror throat, and $\xi_0$ is the mapping between the local $\xi$ and the midplane $\xi_0$ described by:
\begin{equation}
\xi_0^2 = \frac{1}{b}\left[\xi^2 - (1-b) \right].    
\end{equation}
The axial density profile is obtained in a similar fashion:
\begin{eqnarray}
    n_\mathrm{hot}(\psi,b) = 4\pi \mathcal{A}(\psi) \int_0^{v_0}  dv \int_0^{\xi_\mathrm{PT,loc}(b)}d\xi\cr \times  v^2 [f_\mathrm{hot}(v,\xi_0(\xi,b))/\tau_s].
\end{eqnarray}
Normalized profiles of $p_\mathrm{hot}$ and $n_\mathrm{hot}$ are shown in Figure~\ref{fig:egedal_profs}. The total electron density may be constructed similarly to total pressure:
\begin{eqnarray}\label{eq:quasineut}
    Z_\mathrm{eff}(\psi)n_e(\psi,b) = n_\mathrm{hot}(\psi,b)  + \nonumber \\  n_\mathrm{GD}(\psi)(1+f_\mathrm{imp}\langle Z^2_\mathrm{imp}\rangle),    
\end{eqnarray}
where we have assumed there is some gas dynamically confined cold lumped impurity species with $f_\mathrm{imp}= n_\mathrm{imp}/n_\mathrm{GD}$ and average charge $\langle Z_\mathrm{imp}^2\rangle$. It is noteworthy that (\ref{eq:quasineut}) implicitly includes the impact of the mirror's ambipolar potential\cite{Pastukhov1974} despite the potential not appearing explicitly in the MHD equations used to derive the GS equation.

To parameterize the above model, the following must be constrained based on experimental measurements or determined through the non-linear equilibrium reconstruction:
\begin{itemize}
    \item The electron temperature profile $T_e(\psi)$ (under normal circumstances, the electron temperature varies weakly with $b$ according to nonlinear Fokker-Planck calculations using the \texttt{CQL3D-m} code).
    \item The effective charge $Z_\mathrm{eff}(\psi)$ (this  can have significant variation with $b$ due to ambipolar trapping, but this has been excluded here for simplicity). Effective charge can also serve as an enhanced scattering parameter in situations with a substantial gas dynamic cool ion fraction or kinetic instabilities.
    \item The gas dynamic pressure profile $p_\mathrm{GD}(\psi)$.
    \item The gas dynamic density profile $n_\mathrm{GD}(\psi)$.
    \item The hot ions' normalization and radial shape function $\mathcal{A}(\psi)$.
\end{itemize}

Our basis has distinct advantages over previous anisotropic pressure bases used in the literature \cite{Taylor1963,Kesner1985,Quon1985,Bilikmen1997,Taylor2015,Kotelnikov2025,Lindvall2025} (an overview of alternative pressure profile models can be found in Appendix~\ref{ap:p_profs}): it is physically motivated, it includes ion-ion scattering which removes discontinuities that can provoke the mirror instability at low $\beta$ as in Kotelnikov,\cite{Kotelnikov2025} it can self-consistently determine pressure and density allowing measurements of both density and diamagnetic field to be used to constrain equilibrium reconstructions, and because it is parameterized using physical quantities, it can provide deeper insights into plasma conditions. 

To efficiently implement our basis in simulations, a lookup table for $p_\mathrm{hot,(\parallel ,\perp)} (\psi,b)/\mathcal{A}(\psi)$ and $n_\mathrm{hot}(\psi,b)/\mathcal{A}(\psi)$ was constructed for $\theta=45^\circ$ and varying values of $Z_\mathrm{eff} = 1.0$--$3.0$ and $T_e = 0.02$--$1.0$ keV. A fixed value of $R_m$ was used as there was not significant variation in the solution for $R_m = 20$--$100$. A lookup table was preferred to on the fly calculations of $p_\mathrm{hot,(\parallel ,\perp)}$ and $n_\mathrm{hot}$ as such calculations are computationally intensive and would make the equilibrium reconstruction too slow to be used in a routine fashion. In the results presented in this work, 14 terms in the series expansion solution to $f_\mathrm{hot}$ in (\ref{eq:f_egedal}) were retained. Additional terms did not improve the solution quality and caused the moment integrals to converge slowly.

\subsection{Nonlinear equilibrium reconstruction using Bayesian optimization with Gaussian processes}

Bayesian optimization with Gaussian processes (BOGP) is a machine learning technique. In BOGP, a Bayesian optimization uses a surrogate model constructed using a Gaussian process as a prior for an objective function $f(\vec x)$ which it tries to maximize or minimize over some high dimensional set of inputs $\vec x$. Compared to conventional Picard or Newton minimization methods used for nonlinear magnetic equilibrium reconstruction in past work,\cite{Anderson2004} minimization utilizing BOGP can converge faster and is resistant to convergence about local minima and in our realization can provide fast uncertainty quantification. Nonlinear reconstruction allows us to use nonlinear bases for equilibrium parameters, like those described in the previous section, with more physically meaningful fitting parameters. This allows us to glean more information from a limited set of measurements than what would be possible with a linear set of basis functions with little inherent physical meaning. These properties enable efficient nonlinear equilibrium reconstruction to be performed such that better reconstructions can be made with less data in lightly diagnosed machines such as WHAM (and future fusion power plants) provided that an accurate underlying basis is employed. 

In this work we seek to minimize objective function:
\begin{equation}\label{eq:chi2_obj}
    \chi^2 = \sum_{i=1}^N\frac{(M_i - \mathcal{M}_i)^2}{\sigma_{M_i}^2},   
\end{equation}
where $\chi^2$ compares $N$ experimental diagnostic measurements $M$ with some Gaussian experimental uncertainty $\sigma_M$, to synthetic diagnostic measurements based on the equilibrium reconstruction $\mathcal{M}$. As anisotropic plasmas at high-$\beta$ may have points in the optimization bounds which exceed the $\beta$ limit or are susceptible to instabilities, we utilize a refinement on BOGP, Scalable Constrained Bayesian Optimization (SCBO) \cite{Eriksson2021} based on the Trust Region Bayesian Optimization (TuRBO) technique. \cite{Eriksson2019} We used \texttt{BoTorch} \cite{Balandat2020} to create an SCBO implementation nearly identical to that described in Eriksson \textit{et al.}\cite{Eriksson2021} with the exception of the use of Latin-Hypercube sampling \cite{Mckay1979} to obtain the initial points rather than Sobol sampling for greater repeatability and somewhat faster convergence. 

The SCBO technique allows optimizations to be constrained such that if the model fails at a given set of candidate basis function parameters $\vec x$ because $\beta>1$ or if there is a loss of equilibrium due to a firehose (\ref{eq:firehose}) or mirror instability (\ref{eq:mirror}), the optimization can still proceed successfully. SCBO's use of surrogates confined to trust regions, rather than global Gaussian process surrogate models, also allows SCBO to scale to situations in which there are many free profile parameters as it does not fall victim to excessive exploration rather than efficient optimization. Presently, a single-objective function (\ref{eq:chi2_obj}) is used, but SCBO can be extended to a multi-objective optimization \cite{Daulton2022} and the application of multi-objective optimizations to equilibrium reconstruction will be investigated in future work. 

In addition to using the Gaussian process method as a minimization tool, since we use a $\chi^2$ objective function, we can use the Gaussian process surrogate model trained during the minimization to identify the confidence regions in $\chi^2$. This allows us to quantify our reconstructions' uncertainty with very little computational overhead. We identify contours in our surrogate model for $\chi^2(\vec x)$ defining confidence interval:
\begin{equation}
    \chi^2 = \chi^2_\mathrm{min} + \Delta\chi^2_{\sigma_1},
\end{equation}
where $\chi^2_\mathrm{min}$ is the minimum value of $\chi^2$ obtained by the optimization. The joint $\chi^2$ confidence interval is defined by $\Delta \chi^2_{\sigma_1,\mathrm{joint}} = 2\gamma^{-1}(N/2,\alpha_{\sigma_1})/\Gamma(N/2)$, where $\gamma^{-1}$ is the inverse lower incomplete gamma function, $\Gamma$ is the gamma function, and $\alpha_{\sigma_1}=0.683$ sets the confidence interval to correspond to one standard deviation. The marginal $\chi^2$ confidence is similarly defined by $\Delta \chi^2_{\sigma_1,\mathrm{margin}} = 1$. In situations where there is substantial uncertainty in the Gaussian process surrogate model along the $\chi^2 = \chi^2_\mathrm{min} + \Delta\chi^2_{\sigma_1}$ contour, it may be necessary to refine the surrogate model. However, in many cases this is unnecessary as the experimental confidence interval is much larger than the numerical confidence interval in the Gaussian process surrogate.

For the small number of profile parameters used in much of this work, SCBO does not converge to the minimum $\chi^2$ solution appreciably faster than conventional gradient-free methods like Nelder-Mead. However, the SCBO technique we use here is less sensitive to initial guesses and the size of the search region than conventional nonlinear minimization techniques, and provides a surrogate model for $\chi^2$ that enabled fast uncertainty quantification and can potentially be used to check for the uniqueness of equilibrium reconstructions. Furthermore, SCBO methods have been shown to remain fast as dimensionality increases \cite{Eriksson2021,Daulton2022} making them applicable to future equilibrium fitting problems in better diagnosed plasmas where more free parameters are used to prescribe the distribution function and plasma profiles. While we use this fitting technique for nonlinear anisotropic equilibrium reconstruction in the mirror, it should work in other devices as well and could enable more effective high-dimensional nonlinear equilibrium reconstruction with automatic uncertainty quantification in toroidal devices.
 \begin{table}
\caption{\label{tbl:verification} The results from the reconstruction of the simulation data using synthetic measurements for the Maxwellian case in Section~\ref{sec:GDMax}, the hybrid case in Section~\ref{sec:hyb} and the kinetic case in Section~\ref{sec:kinSS}.}
\begin{tabular}{@{}lccccc}
\textbf{Case:} &Max. &Hyb. (1 ms),& (2 ms),& (3 ms) &Kin. \\
\hline
\textbf{Simulated:} \\
\hline
$\langle \beta_0 \rangle$& 0.012&0.066&0.12 &0.15&0.052\\
$\langle E_i \rangle$ [keV]& 0.15&0.88&3.7 &5.9&10.1\\ 
$W_\mathrm{tot}$ [J]& 44& 328&590 &806&263\\
FL1 [$\mu$Wb]& 72& 530& 1000&1400&430\\
FL2 [$\mu$Wb]& 36& 340& 623&860&291\\
FL3 [$\mu$Wb]& 6.8& 51& 90&120&42\\
\hline
\textbf{Reconstructed:} \\
\hline
$\langle \beta_0 \rangle$& 0.011& 0.076&0.15 &0.20&0.065\\
$\langle E_i \rangle$ [keV]&0.13&0.83&3.9 &6.5&12.9 \\
$W_\mathrm{tot}$ [J]& 42& 322&586 &782&266\\
FL1 [$\mu$Wb]& 73& 526& 978&1310&424 \\
FL2 [$\mu$Wb]& 35& 338& 616&854&281 \\
FL3 [$\mu$Wb]& 6.8& 51& 93&124&43 \\
$p_\mathrm{GD,0}$ [Pa]& 133& 1008&1830 &1690&324  \\
$\mathcal{A}_0$ [Pa\,s]& 0& 1167&2030 &3000&1202 \\
$N_\mathrm{fast}/N_\mathrm{tot}$ &0.0&0.05&0.19&0.35&0.7 \\
\hline
\end{tabular}
\end{table}

\subsection{Verification of \texttt{Pleiades} equilibrium reconstructions using numerical data}
\begin{figure*}
    \centering
    \includegraphics{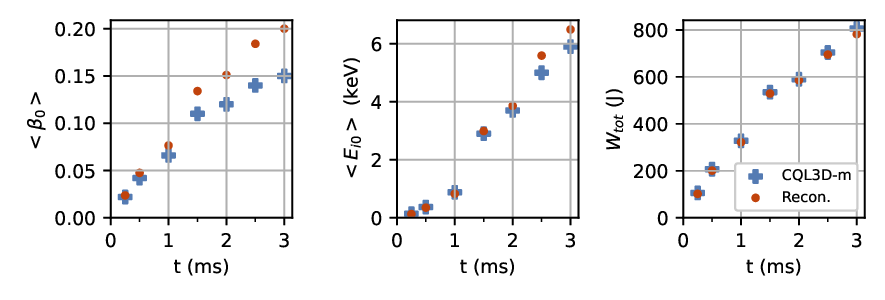}
    \caption{A comparison between \texttt{CQL3D-m}/\texttt{Pleiades} simulation results (blue crosses) and the physics parameters reconstructed from the simulations using synthetic measurements (orange dots) for the ``hybrid'' case described in Section~\ref{sec:hyb}.}
    \label{fig:reconTraces}
    \centering
\end{figure*}
%


To verify the reconstruction technique described here, a reconstruction was performed on synthetic data generated using \texttt{CQL3D-m/Pleiades} simulations similar to those described in recent publications.\cite{Frank2025,Tran2025} These simulations utilized WHAM magnetic field geometry with $B_m = 17.0~\mathrm{T}$ and $B_0 = 0.27~\mathrm{T}$ as well as accurate neutral beam parameters: $E_\mathrm{NBI} = 25~\mathrm{keV}$, $\theta_\mathrm{NBI} = 45^\circ$, and a circular neutral beam source with a $0.2~\mathrm{m}$ diameter, $1^\circ$ divergence, and a $3.4~\mathrm{m}$ focal length, placed $3.8~\mathrm{m}$ from the plasma. 

Synthetic ``measurements'' were taken of the excluded magnetic flux at the flux loop locations in the experiment, $(R,Z) = \{(0.25,0.08),\, (0.2,0.35),\, (0.15,0.62)\}~\mathrm{m}$, as well as electron density and temperature at the midplane at six radial points $R \approx \{ 0.0,\, 0.028,\,0.055, \, 0.084, \, 0.114, \, 0.145\}~\textrm{m}$ at the midplane $Z=0~\mathrm{m}$ (corresponding to the radial views of the Thomson system described in Fujii \textit{et al.}\cite{Fujii2025}). This represents the diagnostic set presently available under nominal WHAM operating conditions. It was assumed in the reconstruction that $T_{e0}(\psi) = T_e(\psi,b)$ (no axial variations in $T_e$) and that $Z_\mathrm{eff}(\psi) = 1.0$ (as in the simulation). In these reconstructions, the $T_e(\psi)$ and $n_e(\psi)$ profiles were set using cubic Hermite interpolation of the synthetic Thomson measurements. The normalized radial shape functions $\mathcal{A}(\psi)/\mathcal{A}_0$ and $p_\mathrm{GD}(\psi)/p_\mathrm{GD,0}$ were assumed to be the same as the normalized Thomson density profile (the normalized electron pressure profile was also considered as a shape function, but the density profile was used as it consistently delivered fits with lower values of $\chi^2$). Each measurement from the synthetic diagnostics was assumed to have no uncertainty ($\sigma = 0$). We used a minimization of modified form of $\chi^2 = \sum (M_i-\mathcal{M}_i)^2/M_i^2$ to find the normalization of the distribution function $\mathcal{A}_0$ on-axis at the midplane and the normalization of the gas dynamic pressure $p_\mathrm{GD,0}$ on-axis at the midplane which provided the best fit to the flux loop measurements. As this is not a true $\chi^2$ test, we could not use the uncertainty quantification method currently employed in the code here, but we plan to develop an enhanced verification suite with realistic noise on measurements in the future. 

Three simulations were reconstructed to verify our reconstruction technique:
\begin{itemize}
    \item Section \ref{sec:GDMax}: A Maxwellian plasma.
    \item Section \ref{sec:hyb}: A time-dependent set of reconstructions for a short NBI pulse into a cold Maxwellian which is shorter than the gas dynamic confinement time creating a hybrid plasma.
    \item Section \ref{sec:kinSS}: A  longer NBI pulse into an initial target plasma intended to generate sloshing ions creating a kinetic plasma.
\end{itemize}
These test cases were designed to determine the reconstruction's ability to: 
\begin{itemize}
    \item Distinguish gas dynamic Maxwellian plasmas from kinetic sloshing ions plasmas.
    \item Accurately measure the total stored energy $W_\mathrm{tot}$. 
    \item Accurately determine the volume averaged values of $\langle\beta_0\rangle$ at the midplane and the ion energy $\langle E_i\rangle$. 
\end{itemize}
A summary of the results of the tests described in the following subsections can be found in Table~\ref{tbl:verification}.

\subsubsection{Gas dynamic Maxwellian} \label{sec:GDMax}
This simulation modeled the WHAM plasma as a uniform Maxwellian plasma with $T = 100~\mathrm{eV}$, $n = 10^{19}~\mathrm{m}^{-3}$, and no NBI. The simulation was initialized and run for a single timestep to generate a plasma which was close to a Maxwellian. The simulation result was then reconstructed. 

Reconstructions correctly identified the plasma as a gas dynamic Maxwellian, but slightly underestimated the ion energy. This is a result of the region around the objective function's minimum being very flat leading to some error in the minimum identification at our level of error tolerance in addition to a mismatch between the \texttt{CQL3D-m} ground truth grids and the reconstruction grids (the reconstruction imposes a Gaussian drop near the limiter that slightly reduces average ion energy). Additional diagnostics would improve convergence, and non-uniform profiles would decrease the error resultant from the edge mismatch. 

\subsubsection{Hybrid kinetic-gas dynamic} \label{sec:hyb}
This simulation modeled a short $P_\mathrm{NBI} = 1~\mathrm{MW}$ NBI pulse into an initially Maxwellian deuterium target WHAM plasma at $t=0$ with $n_e = 5\times10^{19}~\mathrm{m}^{-3}$ and $T_e = T_i = 10~\mathrm{eV}$. The simulation was evolved for $3.0~\mathrm{ms}$ and reconstructions were performed at $0.25~\textrm{ms}$ and $0.5~\textrm{ms}$ then every $0.5~\textrm{ms}$ afterward. The resulting time histories of the simulated values and reconstructed values of key parameters: $\langle\beta_0\rangle$, $\langle E_{i0}\rangle$, and $W_\mathrm{tot}$ are shown in Figure~\ref{fig:reconTraces}.

Reconstructions indicated both gas-dynamic ions heated by NBI slowing down and hot kinetic sloshing NBI ions contributed to the pressure profile. The contribution of fast ions to the overall plasma $N_\mathrm{fast}/N_\mathrm{tot}$ increased over time as anticipated. Initially, agreement of the simulated parameters and reconstructed measurements was precise, however, the quality of the $\chi^2$ fits degraded over time and a $\sim 10\%$ overprediction of $\langle E_{i0} \rangle$ and $\sim 20\%$ overprediction $\langle \beta_0 \rangle$ developed. This may be a result of divergence between the fast-ion basis, which assumes steady-state and no trapped ions at the midplane from sloshing ion induced ambipolar potential trapping, away from the underlying behavior of the ion distribution in \texttt{CQL3D-m} as $T_e$ and the ambipolar potential $\phi$ increase. Another probable source of error are differences in the reconstruction's assumed radial profiles from the actual radial profiles. This could be resolved with a larger number of diagnostics which would allow the profile to be constrained rather than fixed to some assumed radial profile. 

\subsubsection{Kinetic sloshing ions} \label{sec:kinSS}
This simulation modeled a WHAM plasma generated by 200~kW of NBI injection. The simulation was initialized with a 100 eV Maxwellian target and evolved to a steady state over 20 ms, sufficient time for the majority of the Maxwellian ions to leak out. At the final time-step of the simulation, a reconstruction was performed. 

The reconstruction indicated that the plasma was predominantly composed of fast ions, accurately determined the values of $W_\mathrm{tot}$, but, like the reconstructions of the hybrid discharge in the previous section, overestimated $\langle \beta_0\rangle$ and $\langle E_i\rangle$. This overestimation, once again, may be the result of the steady state approximation and the sloshing ion distribution trapping cool plasma at the midplane leading to divergence of the \texttt{CQL3D-m} solution from our kinetic ion basis as well as differences in the assumed and actual radial profiles.

\section{Reconstructing WHAM experiments}\label{sec:exp}
\begin{figure*}
    \centering
    \includegraphics{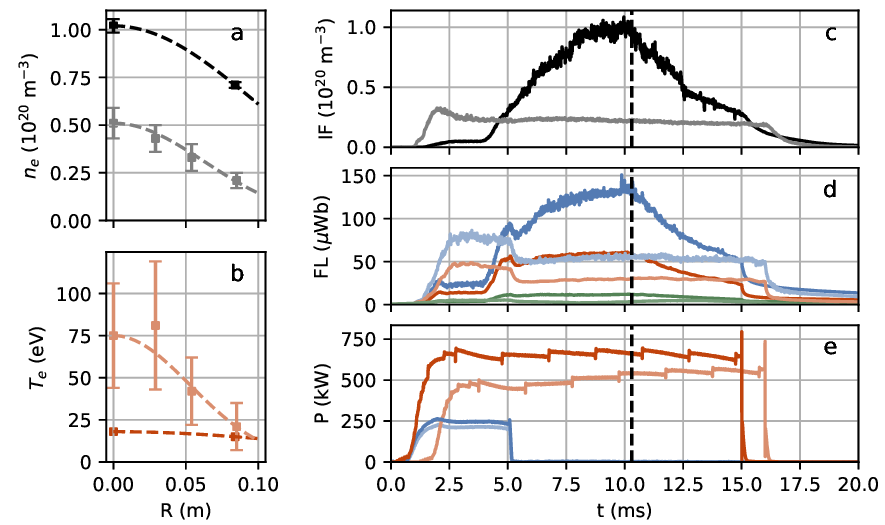}
    \caption{Measurements from the ``high-density'' 250305121--43 (dark) and ``moderate-density'' 250306045--64 (light) WHAM experiments. On the left are the Thomson scattering measurements of $n_e$ in (a) and $T_e$ in (b)  as well as dashed lines showing their Gaussian fits. On the right are the time traces from high-density shot 250305132 (dark) and moderate-density shot 250306062 (light) showing the interferometer density measurement in (c), the flux loop measurements for flux loops 1 (blue), 2 (orange), and 3 (green) in (d), and forward ECH power (blue) and NBI power (orange) in (e). The black dashed vertical line on the right hand side plots denotes 10.3 ms when the Thomson scattering data was collected.}
    \label{fig:fujiiTraces}
\end{figure*}
\begin{table}
\caption{\label{tbl:eqRecon} The results from the reconstruction of the experimental data in Fujii, \textit{et al.}\cite{Fujii2025} including comparisons between the measured flux loop signals and the reconstructed flux loop signals. The Thomson scattering signals used by the reconstruction are shown in Figure~\ref{fig:fujiiTraces}. Note, the error bounds for each reconstructed quantity do not necessarily all occur for the same sets of $p_\mathrm{GD,0}$ and $\mathcal{A}_0$ parameters. These bounds represent the maxima and minima on the joint probability distribution surface.}
\begin{tabular}{@{}lrr}
\textbf{Experiments:} &250305121--43&250306045--64\\
\hline
Measured: \\
\hline
FL1 [$\mu$Wb]& 130$^{(+13/-13)}$& 57$^{(+6/-6)}$\\
FL2 [$\mu$Wb]& 56$^{(+6/-6)}$& 30 $^{(+3/-3)}$\\
FL3 [$\mu$Wb]& 11$^{(+1/-1)}$& 2.5 $^{(+1.3/-1.3)}$\\

\\
Reconstructed: \\
\hline
$p_\mathrm{GD,0}$ [Pa]& 612 $^{(+58/-111)}$  & 333 $^{(+107/-209)}$\\
$\mathcal{A}_0$ [Pa\,s]& 0 $^{(+91/-0)}$ & 131 $^{(+200/-100)}$\\ 
$\langle \beta_0 \rangle$& 0.02 $^{(+0.0015/-0.0015)}$& 0.009 $^{(+0.001/-0.001)}$ \\
$\beta_0(\psi\mathbin{=}0)$& 0.03 $^{(+0.005/-0.002)}$& 0.038 $^{(+0.001/-0.005)}$\\
$\langle E_i \rangle$ [eV]&75 $^{(+10/-11)}$&111 $^{(+20/-20)}$ \\
$W_\mathrm{tot}$ [J]& 71 $^{(+7/-8)}$& 33 $^{(+3/-4)}$\\
$W_\mathrm{fast}/W_\mathrm{tot}$& 0 $^{(+0.2/-0.0)}$&0.17 $^{(+0.35/-0.13)}$ \\
FL1 [$\mu$Wb]& 120 $^{(+13/-12)}$& 53 $^{(+5/-7)}$\\
FL2 [$\mu$Wb]& 60 $^{(+6/-6)}$& 29 $^{(+3/-2)}$\\
FL3 [$\mu$Wb]& 11 $^{(+2/-2)}$& 5.4 $^{(+0.6/-0.7)}$\\

\end{tabular}
\end{table}
\begin{figure}[ht!]
    \centering
    \includegraphics{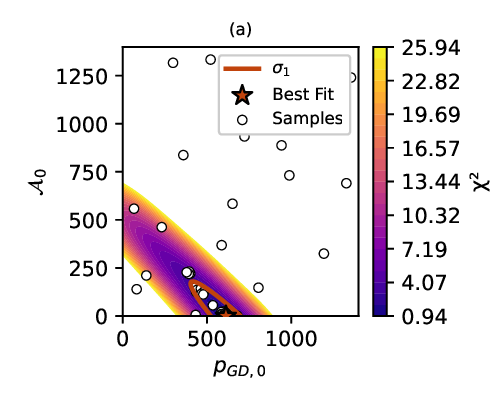}
    \includegraphics{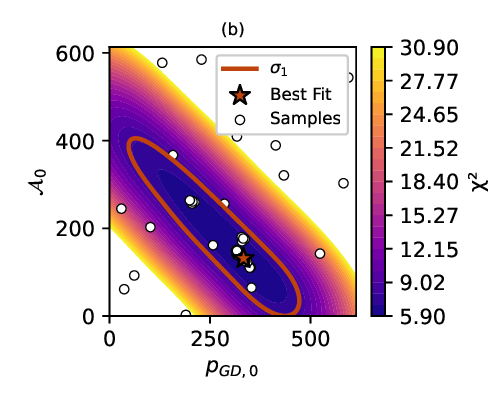}
    \caption{Plots of the surrogate model for $\chi^2$ versus the normalization on the fast ion distribution $\mathcal{A_0}$ and the Maxwellian ion pressure $p_{\mathrm{GD},0}$ for experiments 250305121--43 (a) and 250306045--64 (b). The joint $\sigma_1$ confidence region is shown by the red contour, samples used to construct the model are shown by white dots and the best fit point is shown with a red star. Sloshing ions can be inferred with much greater confidence in the 250306045--64 experiments.}
    \label{fig:chiJoint}
\end{figure}
Two sets of experiments utilizing Thomson scattering were analyzed to determine if the presence of sloshing ions could be inferred in WHAM experiments. These were shots 250305121--43 denoted ``high-density" and shots 250306045--64 denoted ``moderate-density" found in Fujii, \textit{et al.}\cite{Fujii2025} (In these series of shots 250305123, 250305124, 250306046, and 250306050 were excluded as they failed to produce plasmas consistent with the others in their series). Reconstructions were performed at the time of Thomson data collection, $10.3~\mathrm{ms}$, using flux loop measurements that were averaged over the shot series. Measurements from flux loops FL1 and FL2 at $(R,Z) = \{(0.25,0.08),\, (0.2,0.35)\}~\mathrm{m}$ respectively with $\sigma_{M_i} = 0.1 M_i$ and the third flux loop FL3 at $(R,Z) = (0.15,0.62)~\mathrm{m}$ with $\sigma_{M_3}=0.5M_3$ were used in the reconstruction. We used a large $\sigma_{M_3}$ as FL3 is prone to significant errors from mechanical noise. The measurements from the diagnostics in these experiments and the Gaussian fits to the Thomson scattering are shown in Figure~\ref{fig:fujiiTraces}. Following the same process used in the verification simulations in Section~\ref{sec:code}, the normalized radial shape functions $\mathcal{A}(\psi)/\mathcal{A}_0$ and $p_\mathrm{GD}(\psi)/p_\mathrm{GD,0}$ were assumed to be the same as the normalized Gaussian fits to the Thomson density profiles. These reconstructions used a fixed $Z_\mathrm{eff}=2.0$ and $T_e(\psi)$ profiles based on Gaussian fits to the Thomson profiles. We then solved for the normalization on the fast-ion profiles $\mathcal{A}_0$ and the normalization on the gas dynamic pressure $p_\mathrm{GD,0}$ using our equilibrium fitting technique. In the future when more diagnostics are available to constrain the fit and a multiobjective reconstruction algorithm has been implemented, $Z_\mathrm{eff}$ and $T_e$ will be allowed to be free parameters. The approach here was chosen as it helps to ensure that the problem is well determined and it performed better than other options in verification tests like those in the previous section. If we do not fix some of the model's parameters beforehand in this way, we end up in a situation in which we have between 7 and 11 measurements but 14 free parameters (or potentially even more if we chose to use complicated radial shape functions) leaving us with an underdetermined problem. 
\begin{figure}[ht!]
    \centering
    \includegraphics{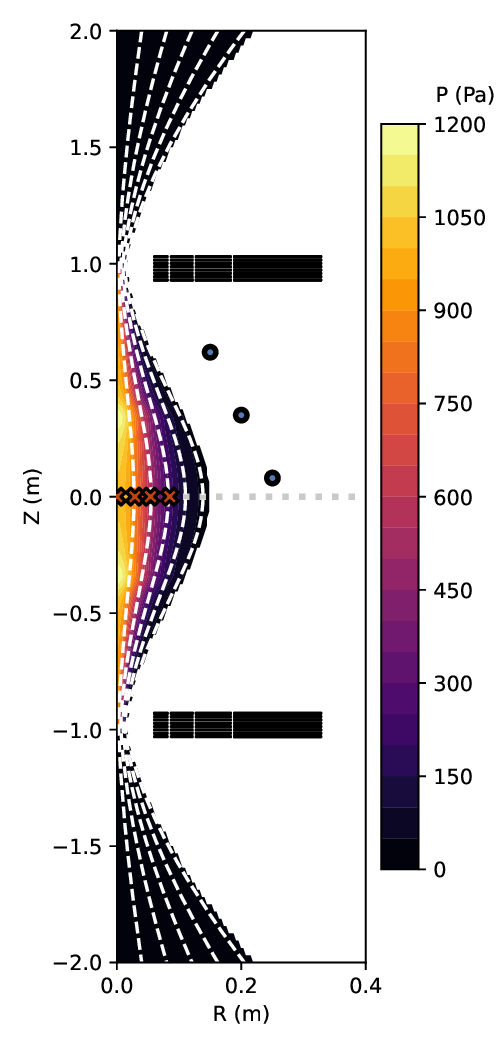}
    \caption{A plot of the reconstructed moderate-density 250306045--64 WHAM equilibrium. Contours of constant $\psi$ are shown by the dashed white lines, the pressure profiles with sloshing ion peaks at $Z\approx\pm0.4~\mathrm{m}$ are shown by the colored contours, the flux loop locations are shown by the blue circles, the Thomson measurement locations are shown by the orange crosses, the interferometer sight-line is shown by the dotted grey line, and the HTS magnets are shown in black at $Z\approx\pm1~\mathrm{m}$.}
    \label{fig:WHAM_slosh}
\end{figure}
\begin{figure}[ht!]
    \centering
    \includegraphics{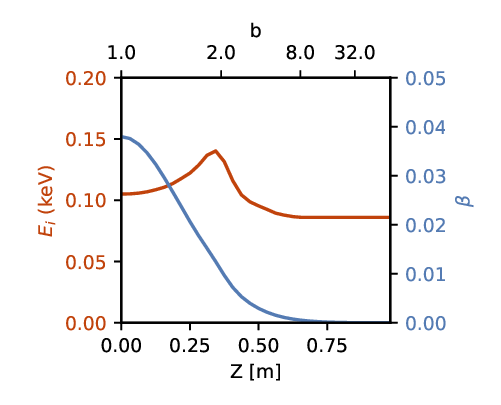}
    \caption{A plot of the on axis ($\psi_n= R = 0.0$) ion energy $\langle E_i \rangle$ and plasma $\beta$ profiles versus the axial location $Z$ and normalized magnetic field $b$ in the reconstructed moderate-density 250306045--64 WHAM equilibrium at the best-fit parameters demonstrating energy peaking from the sloshing ions which induces the pressure peaking seen in Figure~\ref{fig:WHAM_slosh}.}
    \label{fig:WHAM_slosh_cuts}
\end{figure}

We hypothesized that the moderate-density shots with less neutral gas fueling would produce a more kinetic plasma with sloshing ions while the more collisional high-density shots with stronger neutral gas fueling would be nearly totally gas dynamic. This was borne out in the equilibrium reconstruction results shown in Table~\ref{tbl:eqRecon} and the plots of the fits' joint confidence regions in Figure~\ref{fig:chiJoint} that indicate there is likely a sloshing ion population in the moderate-density shots and the high-density shots are likely gas dynamic. Sloshing ion pressure and energy peaks about the fast ion turning point are clearly visible in the contour plots of the reconstructed WHAM equilibrium with the best-fit pressure profiles and flux contours for the moderate-density case shown in Figures~\ref{fig:WHAM_slosh}~and~\ref{fig:WHAM_slosh_cuts}.

A clear confounding factor in this analysis is the potential presence of sloshing fast electrons from ECH. In the moderate and high density experiments investigated here, we did not observe strong hard X-ray signals indicative of fast electron generation, but ideally, an additional basis function for the fast electron contribution to the pressure and density profiles would be included in the reconstruction. However, deriving a semi-analytic distribution function basis for the 2D fast electron distribution under quasilinear diffusion is a more complicated problem than the fast ion basis analyzed here. Additionally, at this time WHAM has insufficient diagnostics to meaningfully constrain three kinetic bases simultaneously without overfitting. A larger number of magnetics measurements, particularly in the regions around the fast electron turning points at $Z\approx\pm(0.7-0.8)~\mathrm{m}$, as well as axially resolved interferometry would improve upon our ability to discern the difference between sloshing ions and electrons in future experiments. We will address both of these issues in future work and WHAM upgrades. However, to make an initial determination as to whether the sloshing ions observed in the moderate-density case were sloshing ions and not sloshing electrons, we devised a test. 

For this test, we compared the quality of fits made with our kinetic pressure profiles (\ref{eq:hot_i_basis}) to those where fast electron pressure profiles were used instead. To model the fast electron pressure profiles, we used the basis in Kesner\cite{Kesner1985} and Quon, \textit{et al.}\cite{Quon1985}:
\begin{subequations}
\begin{eqnarray}
    p_\parallel =& A(\psi)(B/B_\mathrm{turn})[1-(B/B_\mathrm{turn})]^2/2 \\
    p_\perp =& A(\psi)(B/B_\mathrm{turn})^2[1-(B/B_\mathrm{turn})],
\end{eqnarray}
\end{subequations}
here $B_\mathrm{turn} = 4~\mathrm{T}$ (the fundamental resonance location for the $110~\mathrm{GHz}$ ECH system in WHAM \cite{Endrizzi2023}), and $A(\psi)$ is a radial shape factor which we constrained based on Thomson scattering using the same method that was used for the fast ions. This basis is ad-hoc, outside of the fact that it satisfies the parallel equilibrium condition (\ref{eq:forceBalPar}), but when used to reconstruct equilibria in the SM-1 mirror, it produced excellent agreement with the magnetics signals from ECH experiments.\cite{Quon1985} The test found that the kinetic ion basis function produced better fits to the measured magnetics signals. The fast electron basis was a poor fit in this problem. The $\chi^2_\mathrm{min}$ solutions when the fast electron basis was used in combination with a Maxwellian basis had $p_\mathrm{hot} = 0.0$ indicating that the flux loop signals were inconsistent with the presence of sloshing electrons in these plasmas. 

\section{Conclusion and future work}

In this work we have provided a framework for equilibrium reconstruction and the measurement of key plasma parameters such as $\langle \beta \rangle$, $\langle E_i\rangle$, and $W_\mathrm{tot}$ in axisymmetric magnetic mirrors. This represents the first measurements of a non-Maxwellian plasma in WHAM, and the first use of fully kinetic basis functions in magnetic equilibrium reconstructions of fusion experiments. 

We developed a free-boundary anisotropic Grad-Shafranov solver in the \texttt{Pleiades} code and developed a nonlinear equilibrium reconstruction method utilizing basis functions based on solutions to the Fokker-Planck equation \cite{Egedal2022} as well as a machine learning enhanced nonlinear optimization algorithm with uncertainty quantification utilizing SCBO. \cite{Eriksson2021} We verified using synthetic measurements that our reconstructions were able to measure the ion energy, plasma beta, and stored energy to within $\sim20\%$ of their actual values using only diagnostics that are presently available in the WHAM experiment. We then reconstructed WHAM experiments in which Thomson scattering data was available. \cite{Fujii2025} The equilibrium reconstructions indicated the presence of sloshing ions in some WHAM experiments, and we demonstrated, using estimates of the fast electron pressure profiles, that these measurements were unlikely to be the result of sloshing electrons. However, future better diagnosed experiments (this work has motivated the construction of a more comprehensive set of magnetics diagnostics in WHAM) and a more rigorous set of fast electron basis functions will be required to fully quantify the importance of fast electrons in reconstructions of WHAM experiments and other sources of reconstruction error. 

In addition to theoretical and experimental improvements above, several computational improvements will be considered in the future. The MORBO multiobjective optimization using the algorithm in Daulton 2022\cite{Daulton2022} will be implemented in which each diagnostic signal is treated as an individual objective function by the reconstruction algorithm. This approach will generate a pareto front of solutions that will allow even more effective quantification of uncertainty. Reconstructions will also be modified to include the impact of flows, as in some cases with relatively cold plasmas, rotational velocities can become large enough to violate $M^2\ll1$. It is also possible that parallel flows could be relevant outside the core in the mirror expanders where there are low fields and fast outflows. These flows are unlikely to affect core reconstructions but could affect the field line curvature and may be important for calculations of MHD stability. Finally, while WHAM has a large vacuum boundary that allows us to neglect wall currents, future magnetic mirrors are anticipated to have more conformal walls as they can be beneficial for stability. Therefore, the impact of the wall and its 3D structure on flux measurements will need to be considered. More sophisticated equilibrium solvers, like the \texttt{Tokamaker} code\cite{Hansen2024} and 3D MHD solvers, which can account for resistive walls more accurately should be adapted for mirrors. 

\appendix

\section{Derivation of the model for $f_\mathrm{hot}(v,\theta)$}\label{ap:FP_sol} 

\begin{figure}
    \centering
    \includegraphics{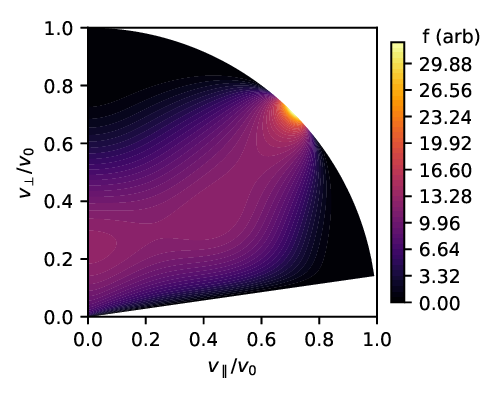}
    \caption{Contours of the hot-ion distribution function $f_\mathrm{hot}$ in arbitrary units for $T_e=500~\mathrm{eV}$, $R_m=50$, $Z_\mathrm{eff}=1.0$, and a neutral beam source with $E_\mathrm{NBI} = 25~\mathrm{keV}$, and $\theta_\mathrm{NBI}= 45^\circ$.}
    \label{fig:egedal_dist}
\end{figure}

The distribution function model from Egedal \textit{et al.} \cite{Egedal2022} is found by solving Fokker-Planck equation:
\begin{equation}\label{eq:fp_orig}
    \frac{\partial f}{\partial  t} = \frac{1}{\tau_s v^2}\frac{\partial}{\partial v}[(v^3 + v_c^3)f] + \frac{Z_\mathrm{eff}}{2\tau_s}\frac{v^3_c}{v^3}\mathcal{L}f + S(v,\theta), 
\end{equation}
where:
\begin{equation}
    \tau_s = \frac{3(2\pi)^{3/2}\epsilon_0^2 T_e^{3/2}m_i}{e^4 Z_i^2 n_e \ln\Lambda_e\sqrt{m_e}},   
\end{equation}
is the slowing down time \cite{Cordey1976}, $\ln \Lambda$ is the Coulomb logarithm, $\epsilon_0$ is the permittivity of free space, $m_i$ and $m_e$ are the ion and electron masses, $Z_i$ is the ion charge, $e$ is the fundamental charge, $T_e$ is the electron temperature, and $n_e$ is the electron density. 
\begin{equation}
    v_c = \left(\frac{3\sqrt{\pi}m_e \ln \Lambda_i}{4 n_e \ln \Lambda_e} \sum_s \frac{Z_s^2 n_s}{m_s} \right)^{1/3}\sqrt{\frac{2 T_e}{m_e}},   
\end{equation}
is the critical velocity (the velocity at which drag on the fast ions by thermal electrons and ions is equal). The Lorentz scattering operator is defined:
\begin{equation}
    \mathcal{L} = \frac{\partial}{\partial \xi} (1-\xi^2)\frac{\partial}{\partial \xi}.   
\end{equation}
For this choice of collisions, the steady state $(\partial f /\partial t \rightarrow 0$) form of equation (\ref{eq:fp_orig}) with a monoenergetic beam source ($S(v,\theta) = \delta(\theta_\mathrm{NBI})\delta(v_0)$) may be solved analytically using separation of variables yielding solution:\cite{Egedal2022}
\begin{equation}
    f(v,\xi) = \frac{\tau_s}{v^3 + v_c^3}\sum_j S_j M_{\lambda_j}(\xi)u^{\lambda_j}    
\end{equation}
where $M_{\lambda_j} = a P_{l_j} + bQ_{l_j}$ is a linear combination of Legendre functions $P_{l_j}$ and $Q_{l_j}$ with non-integer indices $l_j$ and $\lambda_j = l_j(l_j + 1)$ which satisfy boundary conditions:
\begin{subequations}
\begin{eqnarray}
    a P_{l_j}|_{\xi=0} + b Q_{l_j}|_{\xi=0} =&  1 \\
    a \partial_\xi P_{l_j}|_{\xi=0} + b \partial_\xi Q_{l_j}|_{\xi=0} =& 0 \\
    M_{\lambda_j}(\xi_{PT}) =& 0.
\end{eqnarray}
\end{subequations}
The source term is defined:
\begin{equation}
    S_j = \frac{1}{4\pi\alpha_j} \int_0^{\xi_\mathrm{TP}} S_0(\xi)M_{\lambda_j}d\xi,
\end{equation}
where normalization:
\begin{equation}
    \alpha_j = \int_0^{\xi_\mathrm{TP}} M_{\lambda_j}^2 d\xi. 
\end{equation}
Finally, $u$ is defined:
\begin{equation}
    u = \left(\frac{v_0^3 + v_c^3}{v^3+v_c^3} \frac{v^3}{v_0^3} \right)^{\beta_m/3},
\end{equation}
with $\beta_m \approx Z_\mathrm{eff}/ 2$ where $Z_\mathrm{eff}$ is the effective charge. Contours of an example $f$ produced by this model are shown in Figure~\ref{fig:egedal_dist}.

\section{Comparisons of $p_{\parallel,\perp}$ bases}\label{ap:p_profs}
\begin{figure}
    \centering
    \includegraphics[width=\linewidth]{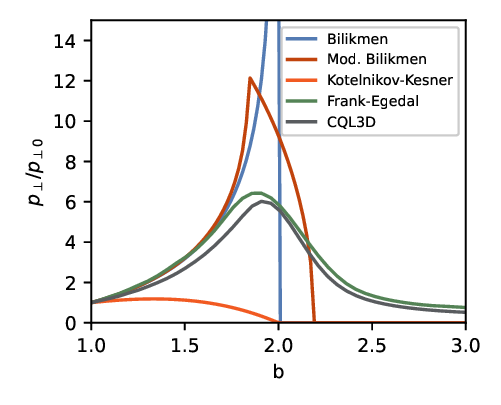}
    \caption{A comparison of normalized $p_\perp$ profiles for 45$^\circ$ NBI injection (i.e. $B_\mathrm{turn} = 2B_0$) including: our ``Frank-Egedal" basis using WHAM parameters with $T_e=50~\mathrm{eV}$, (\ref{eq:kkp}) using $n=1/2$ (Bilikmen) and $n=2$ (Kotelnikov-Kesner), (\ref{eq:bilikmen_mod}) with $\Delta\theta=3^\circ$ (Mod. Bilikmen), and a \texttt{CQL3D-m} WHAM simulation.}
    \label{fig:press_comp}
\end{figure}
Several other models have been used for the axial pressure profiles of sloshing ion distributions in previous studies of mirrors. Of particular note is the model found in numerous mirror studies:\cite{Kesner1985, Quon1985, Bilikmen1997, Kotelnikov2025}
\begin{subequations}
\label{eq:kkp}
\begin{eqnarray}
    p_\parallel =& A(\psi)\bar{b}[1-\bar{b}]^n/n \\
    p_\perp =& A(\psi)\bar{b}^2[1-\bar{b}]^{n-1} ,
\end{eqnarray}
\end{subequations}

where $n$ is some positive constant and $\bar b = b/(B_\mathrm{turn}/B_0) = B/B_\mathrm{turn}$ is the local magnetic field normalized by the magnetic field at the fast ion turning point $B_\mathrm{turn}$. For $B>B_\mathrm{turn}$ it is assumed $p_\perp = p_\parallel = 0$. It can be shown that the steady-state solution to FP equation (\ref{eq:fp_orig}) without the Lorentz scattering term is equivalent to the case in which $n=1/2$ in (\ref{eq:kkp}).\cite{Bilikmen1997} In addition, profiles with $n=2,3$ have been used in stability studies in the past.\cite{Kesner1985, Kotelnikov2025} A difficulty associated with using profiles with $n=2$ or $n=1/2$ is the discontinuity in $\partial p_\perp / \partial B$ at $B=B_\mathrm{turn}$ which can trigger the mirror instability (\ref{eq:mirror}). In practice however, due to finite numerical grid size, the mirror instability does not occur immediately but instead at some finite value of $\beta$ that is much lower than other profiles. \cite{Kotelnikov2025} A profile that somewhat relaxes the singularity in $n=1/2$ case in (\ref{eq:kkp}) can be derived by introducing an NBI source with finite pitch angle width extending from $\theta_l$ to $\theta_u$ i.e.: $S(v,\theta) = \alpha \delta (v-v_0)[\Theta(\theta - \theta_l)\Theta(\theta_u-\theta) + \Theta(\theta - \theta_u - \pi)\Theta(\theta_l + \pi -\theta)]$ where $\Theta(x)$ are Heaviside functions. On integration this yields:
\begin{subequations}
\label{eq:bilikmen_mod}
\begin{eqnarray}
    p_\parallel =& \frac{8}{3}\pi \alpha \tau_d v^2_0 (\varepsilon_0 - \varepsilon_c)\Big[(1-b\sin^2\theta_l)^{3/2} \nonumber \\  &- \Theta(1-b\sin^2\theta_u)(1-b\sin^2\theta_u)^{3/2} \Big] \\
    p_\perp =& \frac{4}{3} \pi \alpha \tau_d v^2_0 (\varepsilon_0 - \varepsilon_c) \Big[(2+ b\sin^2\theta_l) \nonumber \\  &\times \sqrt{1-b\sin^2\theta_l} - \Theta(1-b\sin^2\theta_u) \nonumber \\  &\times (2+b\sin^2\theta_u)\sqrt{1-b\sin^2\theta_u} \Big].
\end{eqnarray}
\end{subequations}
A comparison of the $p_\parallel$ and $p_\perp$ profiles from these models (\ref{eq:kkp}), (\ref{eq:bilikmen_mod}) and the model based on the solution to (\ref{eq:fp_orig}) is shown in Figure~\ref{fig:press_comp}. The basis developed here including scattering noticeably broadens the pressure profile about the turning point and provides a better match to \texttt{CQL3D-m} solutions. Note: we do not expect the solutions provided by the basis function derived here to precisely agree with \texttt{CQL3D-m}. The results produced by \texttt{CQL3D-m} are time dependent, include the ambipolar potential, and utilize a full Rosenbluth collision operator as well a Monte-Carlo neutral beam source model.

\begin{acknowledgments}
This work was funded in part by Realta Fusion and the Advanced Research Projects Agency-Energy (ARPA-E), U.S. Department of Energy under Award Numbers DE-AR0001258, DE-AR0001261, and U.S. Department of Energy under Award Number DE-FG02-ER54744.

This research also used resources of the National Energy Research Scientific Computing Center, a DOE Office of Science User Facility supported by the Office of Science of the U.S. Department of Energy under Contract No. DE-AC02-05CH11231 using NERSC
award FES-ERCAP0026655.
\end{acknowledgments}

\section*{Data Availability Statement}
Datasets associated with the material presented in this paper will be made available on request.

\bibliography{library}

\end{document}